\documentclass[acmlarge]{acmart}

\AtBeginDocument{%
  }

\setcopyright{acmlicensed}
\acmJournal{IMWUT}
\acmYear{2024} \acmVolume{8} \acmNumber{4} \acmArticle{185} \acmMonth{12}\acmDOI{10.1145/3699743}

\acmBooktitle{Proc. ACM Interact. Mob. Wearable Ubiquitous Technol.,
 June 00--00, 0000, Some place, Some state}
\acmISBN{978-1-4503-XXXX-X/18/06}

\usepackage{subfig}
\begin{document}

\title{SleepNetZero: Zero-Burden Zero-Shot Reliable Sleep Staging With Neural Networks Based on Ballistocardiograms}

\author{Shuzhen Li}
\authornote{equal contribution}
\affiliation{%
  \institution{Tsinghua University}
  \streetaddress{30 Shuangqing Rd}
  \city{Beijing}
  \country{China}}
\email{thulisz21@gmail.com}
\orcid{0009-0007-8244-1510}

\author{Yuxin Chen}
\authornotemark[1]
\affiliation{%
  \institution{Tsinghua University}
  \streetaddress{30 Shuangqing Rd}
  \city{Beijing}
  \country{China}}
\email{yuxin-ch21@mails.tsinghua.edu.cn}
\orcid{0009-0006-0477-1357}

\author{Xuesong Chen}
\affiliation{%
  \institution{Beijing Wuji Medical Technology Co., Ltd.}
  \streetaddress{1 Zhongguancun Rd (E)}
  \city{Beijing}
  \country{China}}
\email{chenxuesong1128@163.com}
\orcid{0000-0003-4875-0743}

\author{Ruiyang Gao}
\affiliation{%
  \institution{Beijing Wuji Medical Technology Co., Ltd.}
  \streetaddress{1 Zhongguancun Rd (E)}
  \city{Beijing}
  \country{China}}
\email{railgun@gaoruiyang.cn}
\orcid{0000-0003-2590-5511}

\author{Yupeng Zhang}
\affiliation{%
  \institution{Beijing Wuji Medical Technology Co., Ltd.}
  \streetaddress{1 Zhongguancun Rd (E)}
  \city{Beijing}
  \country{China}}
\email{bookshu1265@gmail.com}
\orcid{0009-0009-7083-3634}

\author{Chao Yu}
\affiliation{%
  \institution{Tsinghua University}
  \streetaddress{30 Shuangqing Rd}
  \city{Beijing}
  \country{China}}
\email{zoeyuchao@gmail.com}
\orcid{0000-0001-6975-0158}

\author{Yunfei Li}
\affiliation{%
  \institution{Tsinghua University}
  \streetaddress{30 Shuangqing Rd}
  \city{Beijing}
  \country{China}}
\email{yunfeili.cloud@gmail.com}
\orcid{0000-0003-0988-9400}

\author{Ziyi Ye}
\affiliation{%
  \institution{Tsinghua University}
  \streetaddress{30 Shuangqing Rd}
  \city{Beijing}
  \country{China}}
\email{yeziyi1998@gmail.com}
\orcid{0000-0002-5622-0235}

\author{Weijun Huang}
\affiliation{%
  \institution{Shanghai Jiao Tong University School of Medicine}
  \streetaddress{227 Chongqing Rd (S)}
  \city{Shanghai}
  \country{China}}
\email{hellohuangwj@126.com}
\orcid{0000-0001-7256-812X}

\author{Hongliang Yi}
\affiliation{%
  \institution{Shanghai Jiao Tong University School of Medicine}
  \streetaddress{227 Chongqing Rd (S)}
  \city{Shanghai}
  \country{China}}
\email{yihongl@126.com}
\orcid{0000-0002-1643-1433}

\author{Yue Leng}
\affiliation{%
  \institution{University of California, San Francisco}
  \streetaddress{675 18th St}
  \city{San Francisco}
  \state{California}
  \country{United States}
  \postcode{94107}}
\email{yue.leng@ucsf.edu}
\orcid{0000-0001-5826-4031}

\author{Yi Wu}
\affiliation{%
  \institution{Tsinghua University}
  \streetaddress{30 Shuangqing Rd}
  \city{Beijing}
  \country{China}}
\email{jxwuyi@gmail.com}
\orcid{0000-0001-9057-5817}

\renewcommand{\shortauthors}{Li et al.}



\received{1 May 2024}
\received[revised]{1 August 2024}
\received[accepted]{20 September 2024}

\begin{abstract}
Sleep monitoring plays a crucial role in maintaining good health, with sleep staging serving as an essential metric in the monitoring process. Traditional methods, utilizing medical sensors like EEG and ECG, can be effective but often present challenges such as unnatural user experience, complex deployment, and high costs. Ballistocardiography~(BCG), a type of piezoelectric sensor signal, offers a non-invasive, user-friendly, and easily deployable alternative for long-term home monitoring.
However, reliable BCG-based sleep staging is challenging due to the limited sleep monitoring data available for BCG. A restricted training dataset prevents the model from generalization across populations. Additionally,
transferring to BCG faces difficulty ensuring model robustness when migrating from other data sources.
To address these issues, we introduce SleepNetZero, a zero-shot learning based approach for sleep staging. To tackle the generalization challenge, we propose a series of BCG feature extraction methods that align BCG components with corresponding respiratory, cardiac, and movement channels in PSG. This allows models 
to be trained on large-scale PSG datasets that are diverse in population. For the migration challenge, we employ data augmentation techniques, significantly enhancing generalizability.
We conducted extensive training and testing on large datasets~(12393 records from 9637 different subjects), achieving an accuracy of 0.803 and a Cohen's Kappa of 0.718. ZeroSleepNet was also deployed in real prototype~(monitoring pads) and tested in actual hospital settings~(265 users), demonstrating an accuracy of 0.697 and a Cohen's Kappa of 0.589.
To the best of our knowledge, this work represents the first known reliable BCG-based sleep staging effort and marks a significant step towards in-home health monitoring.

\end{abstract}

\maketitle

\section{Introduction} \label{sec:intro}

Sleep is a vital physiological process for humans to maintain health and homeostasis~\cite{tobaldini2013heart}.
The consequences of sleep disorders span from daytime sleepiness to increased cardiovascular disease and stroke risk~\cite{roebuck2013review}.
The current gold standard of sleep disorder diagnosis is the \textit{polysomnogram}~(PSG)~\cite{roebuck2013review}.
PSGs are collections of various physiological signals, including \textit{electroencephalograms}~(EEGs), \textit{electrocardiograms}~(ECGs), \textit{electrooculograms}~(EOGs), \textit{electromyograms}~(EMGs), \textit{photoplethysmography}~(PPG), and respiratory signals~\cite{alickovic2018ensemble}.

As a vital aspect of sleep analysis, sleep staging, also known as sleep stage classification, is pivotal for sleep disorder diagnosis~\cite{carskadon2011monitoring,schwartz2008neurophysiology}.
Namely, based on EEG changes, overnight sleep is divided into wake~(W), rapid eye movement~(REM, R) sleep related to dreaming, and three non-REM sleep stages~(N1 -- N3 from light to deep), according to 
the AASM standards~\cite{berry2012aasm}.
Sleep staging helps detect potential health problems, including sleep apnea, stroke, diabetes, brain injury, Parkinson's disease, depression, and Alzheimer's disease~\cite{bianchi2010obstructive, stefani2020sleep, pallayova2010differences, riemann2001sleep, siengsukon2015sleep, mantua2018systematic, zhang2019alteration}.
Traditionally, sleep staging is performed by experienced technicians reading EEGs, assisted by EOGs and EMGs~\cite{berry2012aasm}.
As the process is labor-intensive, automated sleep staging with machine learning methods has been a research focus.

While state-of-the-art sleep staging methods have reached human-level performance~\cite{phan2022sleeptransformer,magalang2013agreement}, they rely on EEGs, whose acquisition requires expensive specialized sensing equipment and is limited in hospital or laboratory environments.
This brings the following drawbacks.
First,
    the extra burden of sleep diagnosis could impede people with sleep disorders from getting diagnosis and treatment, as evidenced by \citet{kapur2002underdiagnosis}, not to mention that longitudinal monitoring is impractical.
Furthermore,
    the subject has to stay in an unfamiliar hospital or laboratory setting, with uncomfortable electrodes stuck to the skin.
    In such environments, the ``first-night effect'' alters natural sleep patterns, thereby impacting the reliability of the results~\cite{herbst2010adaptation,edinger1997sleep}.
Finally,
    setting up the monitoring environment is burdensome for physicians, patients, researchers, and study participants.
Though some EEG-based devices like Dreem~\cite{dreemhealth} and MUSE~\cite{krigolson2021using} can monitor sleep stage changes overnight outside a laboratory setting, they also burden users to sleep as the electrodes have to be attached tightly to their heads.

In contrast, recent studies have utilized household or wearable sensors to circumvent environmental constraints, such as \textit{photoplethysmograms}~(PPGs)~\cite{radha2021deep,li2021transfer,kotzen2022sleepppg} and \textit{ballistocardiograms}~(BCGs)~\cite{migliorini2010automatic,yi2019non,rao2019deepsleep,mitsukura2020sleep,wu2023sleep}.
In particular, BCGs, which ``capture the ballistic forces of the heart caused by the sudden ejection of blood into the great vessels with each heartbeat, breathing, and body movement'', can be acquired with piezoelectric sensors~\cite{sadek2019ballistocardiogram}, which can be deployed easily at home and cause no burden for consumers.

While current BCG-based sleep staging methods show high accuracy, their reliability faces the following challenges.
First,
    annotated BCG data are hard to acquire, as taking BCGs is uncommon in sleep monitoring.
    Although there have been large-scale open-source sleep datasets such as SHHS~\cite{quan1997sleep,zhang2018national}, MESA~\cite{chen2015racial,zhang2018national}, and HSP~\cite{westover2023human} that contain thousands of subjects, all of them do not contain BCG data.
    Therefore, existing methods are trained and tested upon restricted datasets~($\le 25$ subjects), bringing concerns about overfitting.
Second,
    existing works have revealed that sleep patterns significantly shift across ages~\cite{mander2017sleep}, and can be affected by diseases~\cite{abbasi2021comprehensive,stefani2020sleep,holmes2007nature}.
    Therefore, evaluating the methods on a diverse population is crucial for reliability.
Finally,
    while \citet{wu2023sleep} transferred existing models to BCGs in a zero-shot manner, the sleep staging performance evaluated on BCGs was unrevealed.
    Such methods may be problematic, as BCG sensors differ in principle from medical sensors.
    In a household scenario, the sensor may receive multiple disturbances, resulting in a lower-quality signal than PSGs.

To tackle the challenges above, this paper introduces SleepNetZero, a zero-shot framework designed for zero-burden signal generalization. Specifically, we leverage extensive publicly available PSG datasets to address the scarcity of BCG samples. Given the BCG encapsulates the collective effects of heartbeat, breathing, and body movement, we can discern and correlate these components with specific PSG channels. The wealth of PSG data at our disposal facilitates the application of advanced deep learning methodologies, thereby enhancing performance and reliability.

To mitigate the sensor gap, we introduce data augmentation methods that are novel to bio-signals. Our approach augments the input BCG components through amplification and speed perturbation, tailoring the model to accommodate the disturbed signals.

We have conducted comprehensive experiments to demonstrate the superiority and reliability of SleepNetZero. Our model has been thoroughly validated using extensive datasets with 12393 records from 9637 different subjects.
Furthermore, our model has been integrated into a prototype -- a non-intrusive monitoring pad equipped with a BCG sensor -- which has seen widespread adoption in domestic and hospital settings.
SleepNetZero has delivered outstanding results on parallel datasets, including PSG and BCG data, collected from 265 users in a hospital setting.
These results substantiate the efficacy of our methodology.
To our knowledge, this represents the first application of a product utilizing BCG signals for high-precision sleep staging.

To summarize, we make the following major contributions:
\begin{itemize}
    \item \textbf{BCG Component Extraction}: We propose to extract three BCG components to leverage vast PSG datasets, scaling the dataset by two orders of magnitude. We adopt sophisticated inter-beat interval~(IBI) extraction pipelines, indicating the heartbeat component. We find the respiration component through certain frequency bands. We design a novel indicator based on the proposed dynamic energy algorithm, finding the body movement component aligned between EEGs and BCGs. This approach addresses the challenge of limited BCG sample availability.

    \item \textbf{Generalization Enhancement}: To improve the generalizability of SleepNetZero across diverse real-life disturbances, we introduce data augmentation techniques. Specifically, we applied amplification and speed perturbation to the input BCG components to accommodate signals with varying characteristics.
    
    \item \textbf{Neural Network Framework}: We formulate the problem by a sequence labeling task and devise a deep learning model upon the extracted components. The model comprises a ResNet-based feature extractor, a Transformer encoder, and a linear classifier. Finally, a neural network framework named SleepNetZero is constructed, which offers theoretical innovation and demonstrates superior performance in empirical testing.

    \item \textbf{Experiments and Prototype Validation}: We conduct comprehensive experiments on the proposed method, achieving a Kappa of 0.718 and an Accuracy of 0.803, which showcases the reliability of SleepNetZero.
    We also verify the SleepNetZero through a real deployed sleep monitoring prototype, which is zero-burden for the subject. The promising results achieving a Kappa of 0.697 and an Accuracy of 0.589, enhance the confidence in the practicability of SleepNetZero.
\end{itemize}

The rest of our paper is organized as follows. Section~\ref{sec:related} summarizes the related works. Section~\ref{sec:preli} formulates the task. 
Section~\ref{sec:idea} deeply analyzes the challenges of the BCG-based sleep staging task and introduces the basic ideas.
Section~\ref{sec:method} presents the details of SleepNetZero framework design. Section~\ref{sec:exp} describes the experimental setup, and Section~\ref{sec:results} displays the results. We discuss the limitations and future works in Section~\ref{sec:discuss}, and finally, conclude this paper in Section~\ref{sec:conclusion}.

\section{Related Work}\label{sec:related}

Sleep staging can be conducted using various types of devices, each bringing its own set of advantages and limitations.
Specifically, sleep staging can be categorized based on the device type into three main groups: medical sensors, wearable devices, and zero-burden devices. Each category offers distinct methodologies for collecting data and levels of intrusion into the user's life, thus influencing both the accuracy of the sleep stage assessment and its practicality for continuous monitoring.

\subsection{Sleep Staging with Medical Sensors}

With advancing deep learning techniques, EEG-based sleep staging based has progressed rapidly~\cite{jia2020sleepprintnet,qu2020residual,eldele2021attention,eldele2022adast,phan2022sleeptransformer,zhao2022deep,vaezi2023as3}.
\citet{phan2022sleeptransformer} proposed a hierarchical Transformer~\cite{vaswani2017attention} architecture based on single-channel EEGs, resulting in state-of-the-art performance as shown by the experimental results over the SHHS dataset~\cite{quan1997sleep,zhang2018national} containing thousands of participants, which has achieved the human level~\cite{magalang2013agreement}.

While EEG-based methods have seen great success, EEGs are not recorded in common medical settings, which limits the usage of these methods.
Alternative signals are explored for sleep staging, including ECGs~\cite{urtnasan2022deep,zhang2023domain}, heart rate~\cite{fonseca2020automatic,sridhar2020deep,sun2020sleep,zhai2020making,goldammer2022investigation,luo2022hierarchical,ma2023automatic,morokuma2023deep}, respiratory effort~\cite{sun2020sleep}, respiration rate~\cite{goldammer2022investigation,luo2022hierarchical,morokuma2023deep}, body movement~\cite{fonseca2020automatic,zhai2020making,ma2023automatic,morokuma2023deep}, and pulse oximetry~\cite{casal2021automatic}.
While these signals are increasingly available for in-home sleep monitoring, household devices usually differ in principle from medical sensors, and hence signals taken by household devices may not have the high precision as specialized medical sensors.
Therefore, the applicability to household scenarios of the mentioned methods remains unexplored.

\subsection{Sleep Staging with Wearable Devices}

With the prevalence of wearable devices, especially smartwatches, recent works have focused on in-home sleep monitoring through wearable sensors, especially wrist-worn ones.
Typical wrist-worn sensors include \textit{photoplethysmography}~(PPG) and actigraphy~(also known as accelerometry).
Methods based on these sensors have been making rapid progress~\cite{beattie2017estimation,li2021transfer,radha2021deep,zhai2021ubi,kotzen2022sleepppg}.
However, tightly worn sensors during sleep may cause discomfort to the skin, while loosely worn sensors may be affected by large artifacts. This impedes the practicality and reliability of the methods based on wearable devices.

\subsection{Sleep Staging with Zero-Burden Devices}

Unlike wearable devices, zero-burden devices are placed in the bedroom, contactless with the subject, further improving the comfort for consumers. Sleep staging through a variety of zero-burden devices has been studied. \citet{zhao2017learning} leveraged radio frequency signals and explored an adversarial architecture of neural networks. \citet{hong2022end} leveraged nocturnal sounds with a neural network trained on the cleaner PSG audio dataset and tested on the noisier smartphone audio dataset. \citet{van2023contactless} utilized cameras above the bed for heartbeat extract, which the authors call ``remote PPGs''.

However, the above methods are remote, whose signals can be easily disturbed by the environment, hindering the robustness. Conversely, physical connections are more reliable. The ballistic force of vessels can be conducted through the pillow and captured by the piezoelectric materials below, validating the superiority of BCGs as it is physical yet zero-burden.

There have been studies on BCG-based sleep staging~\cite{migliorini2010automatic,yi2019non,rao2019deepsleep,mitsukura2020sleep}. Due to the lack of data, \citet{migliorini2010automatic,yi2019non,mitsukura2020sleep} utilized traditional machine learning methods. \citet{rao2019deepsleep} proposed a deep learning method that addressed the data shortage by pre-training with a mixture of raw ECGs, PPGs, and BCGs. All these works conducted experiments on no more than 25 subjects.

While pre-training could potentially leverage the PSG dataset, the mechanism is quite opaque. Moreover, transfer learning splits the BCG dataset for training and validation, reducing the test dataset. In contrast, our proposed method leverages the vast PSG dataset by component extraction, which requires no transfer learning. As a result, the precious BCG dataset could be held out entirely for testing, significantly improving the reliability of the experimental results.

Resembling our method, \citet{wu2023sleep} did propose a method that constructs ECGs from BCG-based features, but the experiments on sleep staging were not conducted upon annotated BCG datasets. Therefore, as far as we know, we are the first to perform the component extraction method on BCG-based sleep staging.

\section{Preliminaries of Sleep Staging}\label{sec:preli}

Sleep staging, as defined by the American Academy of Sleep Medicine~(AASM) standards~\cite{berry2012aasm}, is a critical process in sleep analysis that involves segmenting the entire duration of overnight sleep into distinct intervals. These intervals are then classified into one of five sleep stages by trained technicians. This categorization is fundamental for understanding sleep patterns and diagnosing sleep disorders.

\subsection{Segmentation and Labeling}

The sleep record for an individual night is divided into $T$ successive segments, where each segment corresponds to a 30-second interval. This 30-second duration is chosen based on AASM guidelines, which recommend it as the standard epoch length for clinical sleep stage scoring due to its effectiveness in capturing sufficient cycles of sleep activity while maintaining manageable data granularity.

For every segment $t$, where $t = 1, 2, \ldots, T$, it represents the time from $30(t-1)$ seconds to $30t$ seconds of the overnight recording. The corresponding sleep stage for each of these segments is denoted by $y_t$, which can take one of the five possible categorical values: W, N1, N2, N3, and R. Their meanings are described in Appendix~\ref{sec:stages}.

\subsection{Data Handling for Short Segments}
In cases where the remaining portion of the sleep record does not neatly fit into a 30-second segment — typically the final few seconds of the recording — this remainder is excluded from the analysis. This practice ensures that every segment analyzed maintains uniform length and data integrity.

\section{Challenges and Basic Idea of SleepNetZero}\label{sec:idea}
In this section, we highlight the aforementioned challenges in Section. \ref{sec:intro}  and present basic ideas to address the challenges.

\subsection{Challenges}
To the end of achieving accurate sleep staging, several challenges need to be solved as follows.
\begin{enumerate}
    \item{\textbf{Diversity of Population}}: Although recent BCG-based sleep staging methods~\cite{yi2019non,migliorini2010automatic,rao2019deepsleep,mitsukura2020sleep} have reported high accuracy, their main defects are the reliability across populations.
    Some works trained and tested the model on data from the same subject~\cite{yi2019non}, which leads to dependence on individual features.
    Others used the leave-one-subject-out method~\cite{migliorini2010automatic,rao2019deepsleep,mitsukura2020sleep}, where the test data is directly used for hyperparameter tuning.
    Possible flaws of such methods include overfitting the test population.
    Therefore, their performance relies on restricted datasets~(no more than 25 subjects).
    As sleep pattern changes across individual features like age or health status, the previous methods may suffer from performance drops when facing unseen data.

    \item{\textbf{Generalization Across Sensors}}: In principle, each PSG channel is taken by a specialized medical sensor, while all BCG components share a household sensor.
    The sensor gap implies that the components extracted from BCGs may have a lower quality than those from PSGs.
    As a result, a model trained with high-quality PSG-derived data may fail to generalize to low-quality BCG-derived data.
    Therefore, the second difficulty arises: we must handle the quality gap between PSGs and BCGs, which is unprecedented.

    \item{\textbf{Multimodal Bio-Signal Modeling}}: Different BCG components are in different modalities. Heartbeats are sparse yet uniformly spaced, breath is continuous, and body movement is a sequence of 0/1 indicators.
    The modeling and fusion of such modalities, especially body movement, lack reference.
    
\end{enumerate}
We present our basic idea to the aforementioned challenges in the following subsections.

\subsection{Basic Idea}
\subsubsection{Leveraging PSG Datasets}
To address the first challenge, we noticed the large-scale open-source PSG datasets, such as SHHS~\cite{quan1997sleep,zhang2018national}, which contains about 5.8k subjects.
The reliability inherently comes from the patients' diversity of ages and diseases.

However, formerly people tend to consider PSG and BCG as separate signals. There are very few works using PSG data to assist in BCG-based sleep staging. Although \citet{wu2023sleep} proposed a pipeline leveraging ECGs from PSG datasets, its performance on annotated BCG data remains unexamined. We are the first to utilize PSG data for BCG-based sleep staging with convincing results.

As mentioned in \citet{sadek2019ballistocardiogram}, BCGs represent heartbeat, breath, and body movement.
Therefore, we aim to extract these components from BCGs and PSGs, bridging both modalities.
For the heartbeat, as suggested by the recent practice~\cite{goldammer2022investigation,ma2023automatic,morokuma2023deep}, we extract inter-beat intervals~(IBIs) and then interpolate them into a continuously varying signal.
The interpolated IBI signal represents the heart rate variability, and the identification pipelines are sophisticated~\cite{makowski2021neurokit2,massaro2019heart,bruser2013robust}.
We propose new methods for the breath and movement components.

\subsubsection{Generalization Across Sensors}

The basic idea to handle the difficulty is to train the model with lower-quality data.
Although the variation from PSGs to the BCG components is unstudied, we may hypothesize some typical perturbation patterns.
As bio-signals are 1-dimensional real-valued signals, we incorporate random amplification and speed perturbation referring to the audio practice.

\subsubsection{Multimodal Bio-Signal Modeling}

The key insight of modeling the multimodal bio-signals comes from the fact that the body movement indicator has only 0/1 values and plays the role of mask. Therefore, we will first introduce the framework of modeling the heartbeat and breath waves, and then adhere the body movement to them.

The common practice to model a single waveform is a hierarchical solution~\cite{baevski2020wav2vec,hsu2021hubert}. First, the high-resolution input is encoded by a convolutional feature extractor, resulting in a local representation for each window. Consequently, the representations are sent into a Transformer encoder~\cite{vaswani2017attention}, capturing contextual information.

Next, we fuse the two modalities. Besides the individual representation of each modality, their coupling should also be captured.
Therefore, we use three distinct feature extractors. The first encodes the heartbeat, the second encodes the breath, and the third encodes both.

Finally, we add the body movement to each feature extractor. Namely, the first feature extractor encodes the heartbeat and body movement, the second encodes the breath and body movement, and the third encodes the three components.

The extracted features are concatenated to form the local representation. The local representations are then sent into the Transformer encoder, giving the contextual information. Finally, the contextual information is sent into a multi-layer perceptron~(MLP), whose output is the probability over the classes.

In conclusion, we use a neural network to model the problem. The neural network comprises three feature extractors, a Transformer encoder~\cite{vaswani2017attention}, and a MLP classifier.

\section{SleepNetZero Framework Design}\label{sec:method}
This section presents the framework of SleepNetZero.
First, an overview of the framework of SleepNetZero is given. 
Then, each module of the framework is described in the following subsections.

\subsection{Overview of SleepNetZero}
\begin{figure}
    \centering
    \includegraphics[width=0.99\linewidth]{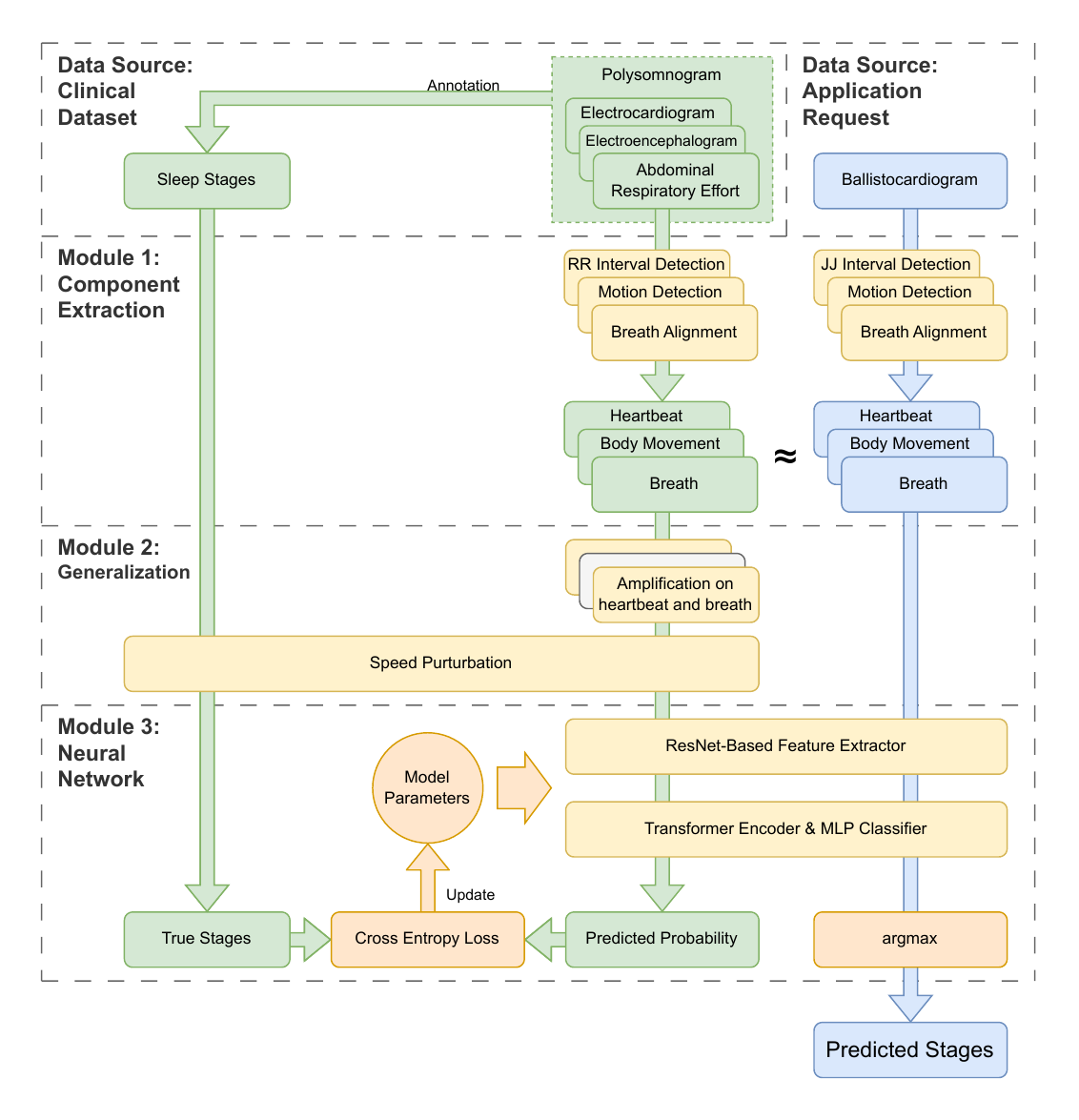}
    \caption{The overview of the proposed framework. Training data from clinical datasets and test data from application requests are aligned through the component extraction module, which comprises three extractors. For training data, the extracted components are augmented by the generalization module, which consists of amplification and speed perturbation. Finally, these components are fed into a neural network for training and inference.}
    \label{fig:method-overview}
    \Description{The three modules (component extraction, generalization, and the neural network) act in a sequence to the input data. The input data is either annotated polysomnograms from clinical datasets or ballistocardiograms from application requests.}
\end{figure}

As shown in Figure~\ref{fig:method-overview}, the proposed framework consists of three main stages: component extraction, generalization, and the neural network.
In the component extraction stage, the framework processes signals from BCG signals to extract meaningful features related to heartbeats, breathing, and body movements. Each type of signal is treated with specialized algorithms to ensure the accuracy and relevance of the data for sleep staging.
Following extraction, the generalization stage prepares the data for deep learning applications with two innovative data augmentation techniques.
The final stage employs a neural network, specifically ResNet architectures and Transformer encoders, to classify sleep stages based on the processed signals. This multi-layered approach leverages deep learning techniques to interpret the intricate patterns in the data, ultimately classifying sleep stages with high precision.

The specific settings and methodologies applied in the model training process are discussed in subsequent sections, detailing the technical strategies that enhance the framework’s performance.

\subsection{Component Extraction Module}

We extract three components representing heartbeat, breath, and body movement. The algorithms are listed below.
\paragraph{Heartbeat Component}

Following the recent practice, we represent heartbeats by inter-beat intervals~(IBIs). The IBI representation enables a lower sampling frequency and offers a convenient artifact removal method. For ECGs, the heartbeats are identified as R peaks, and the IBI is known as the RR interval~(RRI). For BCGs, the heartbeats are identified as J peaks, and the IBI is known as the JJ interval~(JJI).

\begin{itemize}
    \item Let $n$ denote the number of heartbeat intervals.
    \item Let $P_i$ denote the end time of the $i^{\rm th}$ heartbeat interval.
    \item Let $T_i$ denote the length of the $i^{\rm th}$ heartbeat interval.
\end{itemize}
From BCGs, we adopt the method proposed in \cite{bruser2013robust} to extract $(P_i, T_i)$.
From PSGs, we clean the ECG and find the R peaks $P_0, P_1, \ldots, P_n$ using NeuroKit 2~\cite{makowski2021neurokit2}, and then assign $T_i=P_i-P_{i-1}$.

To prevent outliers, while preserving the end timestamps $P_i$, we replace the intervals $T_i$ with normal-to-normal intervals using the \texttt{get\_nn\_intervals} function in the package HRVAnalysis~\cite{massaro2019heart} with default parameters, i.e., $T_i \leftarrow \texttt{get\_nn\_intervals}(T_i)$. The function removes physiologically implausible or ectopic $T_i$'s. The removed values are linearly interpolated between the remaining ones. Removed $T_i$'s at the beginning or the end of the sequence are dropped, and the corresponding $P_i$'s are dropped, too.
Note that $P_i$ does not necessarily equal $P_{i-1}+T_i$ after the correction.

The heartbeat signal fed into the model is linearly interpolated at a sampling frequency of $4\rm Hz$ between the anchor points $(P_i, T_i)$. The signal before $P_1$ is padded by the constant $T_1$, and the signal after $P_n$ is padded by the constant $T_n$.
\paragraph{Breath Component}

The breath signal resides in the BCG and the abdominal respiratory effort~(ABD effort) in the PSG. The model input is extracted through our proposed four-stage pipeline: filtering, resampling, integration, and normalization.

First, a $\dfrac{1}{10} \sim \dfrac{1}{3} \mathrm{Hz}$ band-pass third-order Bessel filter~\cite{thomson1949delay} is applied to the raw BCG signal and the ABD effort.
The output is then resampled to a sample frequency of $4\rm Hz$. Next, we integrate the BCG-derived breath signal over time. The integration is not applied to the PSG-derived breath. Finally, the final signal is normalized to z-scores.

The integration remedies the phase gap between BCG signals and ABD effort, since the BCG signal represents the breathing flux, while the ABD effort reflects the volume of the breathed air.
The normalization removes the magnitude difference, which is because the two types of sensors differ in principle.

\paragraph{Body Movement Component}

Body movements appear as significant disturbances to EEGs~\cite{berry2012aasm} and BCGs.
As BCG and EEG data is sampled at a rate of 500Hz, given the frequent changes in the BCG signal during body movements, a 2-second window is appropriate to capture these rapid variations. Within this 2-second window, we use the peak-to-peak amplitude as a feature to characterize the signal changes.

For each window, we establish a 30-second dynamic baseline of statistical metrics, including the mean~($\mu$) and standard deviation~($\sigma$), calculated from the windowed data. The baseline period is divided into 15 2-second windows. Each of these 2-second windows provides a feature value, forming a sequence of 15 values. We then compute the mean and variance of this sequence. This baseline helps in adapting to the intrinsic variability of the EEG or BCG signals over time. A threshold is then applied to detect significant deviations from this baseline, indicative of potential body movements. Specifically, any window where the peak-to-peak amplitude exceeds $\mu + 5\sigma$ is flagged as containing body movement. This threshold is chosen based on the distribution characteristics of the signal and is intended to ensure that only substantial deviations, which are likely due to movements rather than physiological fluctuations, are considered. The use of a multiplier~(5) of the standard deviation is guided by statistical norms where such a level typically signifies a departure from normal variance, assuming a Gaussian distribution.

\subsection{Generalization Module}

The Generalization Module is designed to enhance the robustness of our model by addressing discrepancies caused by sensor gaps. We introduce two innovative data augmentation techniques specifically tailored for bio-signals. These techniques are random amplification and speed perturbation, each targeting different aspects of signal variation. 
Namely, the random amplification stretches the signal in the magnitude direction, and the speed perturbation stretches the signal in the time direction.

\subsubsection{Random Amplification}
The first technique, random amplification,{\color{black} aims to simulate variations in signal intensity that might occur due to sensor sensitivity or user differences}. This technique adjusts the amplitude of the heartbeat and breath signals independently, making the model more resilient to fluctuations in signal strength.
The heartbeat signal and the breath signal are amplified independently.
For the signal $\bf x$, we pick a scalar $\alpha \sim \mathrm{Uniform}(0.9, 1.1)$, and the model input is replaced by $\alpha\bf x$.
This process ensures that the model can handle slight variations in amplitude without compromising the accuracy of stage classification.

\subsubsection{Speed Perturbation}
The second technique, speed perturbation, addresses variations in the temporal domain, mimicking scenarios where the signal sampling speed might vary due to technical issues or user differences.
This technique uniformly stretches or compresses the three input components of the model along with their corresponding ground truth stages.

The speed perturbation affects the three components and the ground truth stages by a common factor $\beta \sim \mathrm{Uniform}(0.75, 1.25)$. The three input components are perturbed adopting the implementation from torchaudio~\cite{yang2021torchaudio,hwang2023torchaudio}.
Once the signal is adjusted, suppose the signal length is $L'$ seconds, then the ground truth should have a length of $T'=\lfloor L'/30 \rfloor$, where 30s is the annotation segment length. The stages $y'$ are transformed by a linear spacing sampling:
\begin{displaymath}
t(i)=\left\lfloor \frac{T(i-0.5)}{T'} \right\rfloor + 1,\ y'_i = y_{t(i)},\ i=1, 2, \ldots, T'.
\end{displaymath}
Intuitively, for each 30-second segment, we find the center's original position and sample the original stage. After that, the signals are truncated to a length of $30T'$ seconds, maintaining a standardized input size for the model.

\subsubsection{Visualization of the Module}
\begin{figure}
    \centering
    \includegraphics[width=0.7\linewidth]{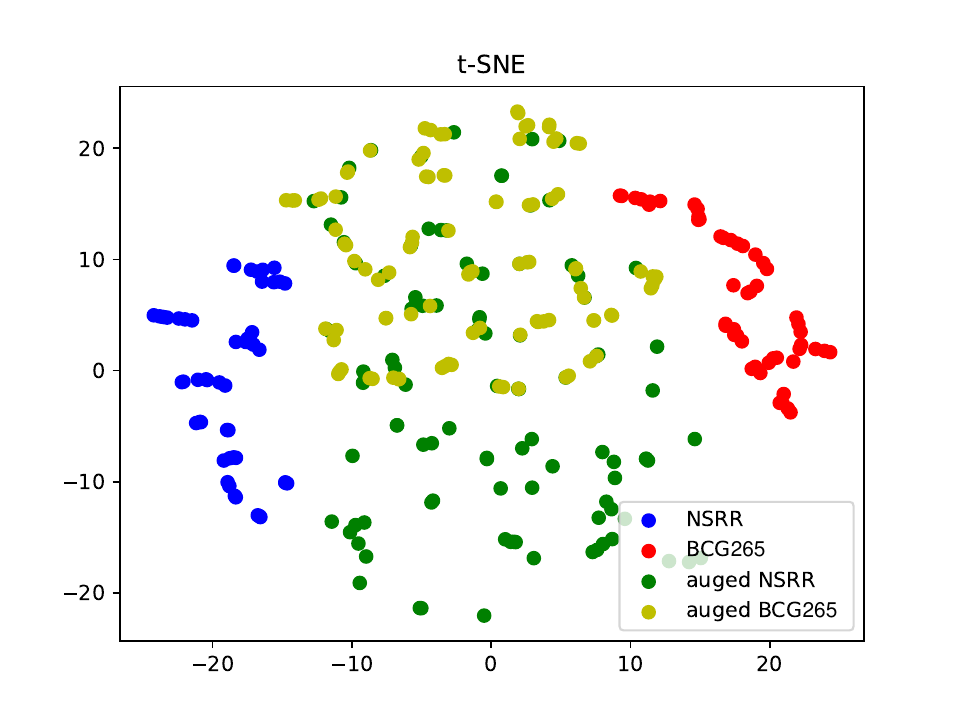}
    \caption{The t-SNE visualization of the proposed data augmentation technique in the spectrum space. The original distributions of NSRR (blue) and BCG265 (red) data are distinct and separate. They cluster closer to each other (green, yellow) after data augmentation, illustrating that our data augmentation effectively eliminates the disparity between the two datasets.}
    \label{fig:tsne}
    \Description{The data from the NSRR dataset concatenate on the left of the figure, while the data from the BCG265 dataset concatenate on the right of the figure. After data augmentation, data from both datasets move toward the center.}
\end{figure}

We demonstrate the effectiveness of the proposed data augmentation technique with Figure~\ref{fig:tsne}. In the demonstration, we randomly pick 30 records from NSRR and BCG265 respectively. The second hour of the record is cropped. Each crop is augmented with three different random seeds. For each crop, we concatenate the Welch power spectrum densities (PSDs) of the heartbeat and breath components to form a feature vector. Then, we visualize the 2D projection with t-SNE of the vectors from various datasets. The data from NSRR and BCG265, originally distributed in distinct and separate areas, cluster closer to each other after data augmentation. This provides a piece of evidence that our data augmentation effectively eliminates the disparity between the two datasets.

\subsection{Neural Network Module}
In the proposed model, we address the task of sleep staging as a sequence labeling problem, utilizing the rich temporal dynamics of physiological signals.

\subsubsection{Data Representation}

The input to our model consists of three distinct physiological signal sequences, denoted as ${\bf x}_1, {\bf x}_2, {\bf x}_3$. ${\bf x}_1, {\bf x}_2$ are continuous signals: ${\bf x}_1$ is the IBI wave representing the heartbeat, and ${\bf x}_2$ is the breath wave. ${\bf x}_3$ is the 0/1 sequence indicating body movements. 
To standardize the input, we select a unified sampling rate of $4$ Hz for each sequence, resulting in sequence lengths of $120T$ for a total recording time of $30T$ seconds.

\subsubsection{Feature Extraction}

\begin{figure}
\centering
\includegraphics[width=0.9\linewidth]{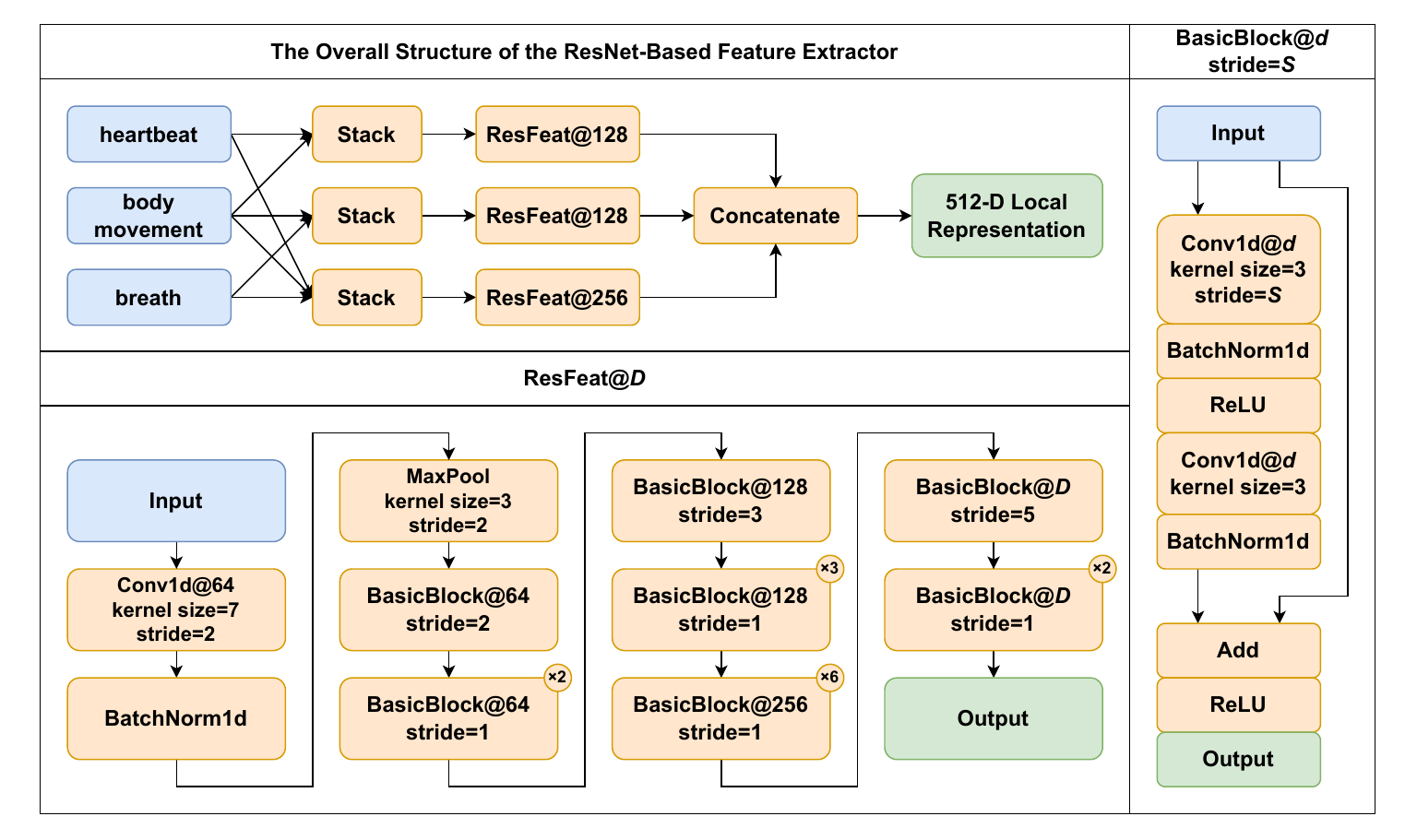}
\caption{This is the architecture of the proposed three feature extractors, key components that extract high-dimensional features and fuse different modalities. The upper left part shows the overall architecture. The three components -- heartbeat, body movement, and breath -- are taken as inputs. Three ResNet-based feature extractors, denoted by ResFeat in the figure, produce high-dimensional representations for the three distinct groupings of components. The representations provided by the three extractors are concatenated together. The lower left part shows the detailed architecture of the ResFeat module, which is the main body of the feature extractor, denoted $\mathbf f_1, \mathbf f_2, \mathbf f_3$ in the main text. The right part shows the detailed architecture of the BasicBlock module, which is part of the ResFeat module.}
\label{fig:resnet}
\Description{
The three ResNet-based feature extractors f_1, f_2, and f_3 has output dimensionalities 128, 128, and 256, respectively.
Each feature extractor consists of multiple blocks.
The first is a convolution layer followed by a batch normalization layer and then a max pooling layer. The convolution layer is 64-dimensional with kernel size 7 and stride 2. The max pooling layer has a kernel size of 3 and a stride of 2.
After the first block, there are 16 basic blocks. The first 3 blocks are 64-dimensional, the next 4 are 128-dimensional, the following 6 are 256 dimensional, and the last 3 match the output dimensionality.
The 1st, 4th, and 14th basic blocks has a stride of 2, 3, and 5, respectively. Others has a stride of 1.
Basic blocks have the same architecture as that in ResNet34, except that all 2-dimensional operators are replaced by their 1-dimensional counterparts.
}
\end{figure}
To capture the unique characteristics of each signal component effectively, we employ three independent feature extractors based on the ResNet architecture~\cite{he2016deep}, as shown in Figure~\ref{fig:resnet}. Each extractor, denoted $\mathbf{f}_1, \mathbf{f}_2, \mathbf{f}_3$, is trained to transform its respective input sequence into a high-dimensional feature space. For each $t^{\rm th}$ classification segment, the features are aggregated as follows:
\begin{equation}
\mathbf{z}_t=\operatorname{concatenate}\Big(\mathbf{f}_1\big(\operatorname{stack}({\bf x}_1^{(t)}, {\bf x}_3^{(t)})\big), \mathbf{f}_2\big(\operatorname{stack}({\bf x}_2^{(t)}, {\bf x}_3^{(t)})\big), \mathbf{f}_3\big(\operatorname{stack}({\bf x}_1^{(t)}, {\bf x}_2^{(t)}, {\bf x}_3^{(t)})\big)\Big),
\end{equation}
where ${\bf x}_i^{(t)}$ denotes the perception field of the $i^{\rm th}$ sequence at time $t$. This design allows the network to learn from the local context of each signal, enhancing the model's ability to discern subtle patterns indicative of different sleep stages.
\subsubsection{Context Encoding and Classification}

Following feature extraction, the aggregated feature vectors $\mathbf{z}_t$ are processed by a Transformer encoder~\cite{vaswani2017attention}, which incorporates a sinusoidal positional encoding~\cite{vaswani2017attention} to maintain the temporal context of the sequence:
\begin{equation}
[\mathbf{c}_1, \mathbf{c}_2, \ldots, \mathbf{c}_T]=\operatorname{encoder}([\mathbf{z}_1+\mathbf{e}_1, \mathbf{z}_2+\mathbf{e}_2, \ldots, \mathbf{z}_T+\mathbf{e}_T]; \theta),
\end{equation}
where $\mathbf{e}_t$ represents the sinusoidal positional encoding for the $t^{\rm th}$ segment. This step is crucial as it allows the model to leverage both the local features extracted by ResNet and the global dependencies between segments captured by the Transformer architecture.

The context vectors $\mathbf{c}_t$ output from the Transformer encoder are then fed into a 2-layer MLP classifier, which computes the probability distribution over sleep stages for each segment using a softmax function:
\begin{equation}
\mathbf{p}_t=\operatorname{softmax}(\mathbf{W_2}\max(\mathbf{0}, \mathbf{W_1}\mathbf{c}_t+\mathbf{b_1})+\mathbf{b_2}),
\end{equation}
where $\mathbf{p}_{t, c}$ indicates the predicted probability of the $t^{\rm th}$ segment being in the $c^{\rm th}$ sleep stage. The final prediction $\hat{y}_t$ for each segment is determined by selecting the class with the highest probability:
\begin{equation}
\hat{y}_t=\arg\max_c \mathbf{p}_{t, c}.
\end{equation}

The Transformer encoder used has 6 layers. Each layer is 512-dimensional, with a 2048-dimensional feedforward neural network. The hidden layer of the MLP classifier is 64-dimensional.

\subsubsection{Training}

During the training phase, we optimize the parameters of the model by minimizing the cross-entropy loss between the predicted probabilities and the true class labels:
\begin{equation}
L=-\sum_{t=1}^T \log \mathbf{p}_{t, y_t},
\end{equation}
where $y_t$ denotes the true class label for the $t^{\rm th}$ segment. This loss function encourages the model to accurately predict the correct sleep stage, thus improving its overall performance on unseen data.

\section{Experimental Setup}\label{sec:exp}
This section outlines the specific configurations of our experiments, including the datasets used, the evaluation metrics applied, and the details concerning the training process.

\subsection{Datasets}
\paragraph{The NSRR Dataset}
We train and evaluate our method using the combination of the following publicly available PSG datasets: SHHS~\cite{quan1997sleep,zhang2018national}, MrOS~\cite{blackwell2011associations,zhang2018national}, MESA~\cite{chen2015racial,zhang2018national}, CCSHS~\cite{rosen2003prevalence,zhang2018national}, CFS~\cite{redline1995familial,zhang2018national}, and HomePAP~\cite{rosen2012multisite,zhang2018national}. We will refer to the combined dataset by NSRR. After cleaning, the NSRR dataset contains 12393 records from 9637 different subjects.
We include its age and sex distributions in Appendix~\ref{app:data}.

Among all the subjects in the NSRR dataset, $10\%$ are held out for testing, $10\%$ are split for validation, and the remaining $80\%$ are used for training the model. The training/validation/test subsets consist of the data from their corresponding subjects. Therefore, recordings from different splits are from distinct subjects. Across experiments, the training/validation/test split does not change.

The ground truth labels are imbalanced. In the NSRR dataset, the distribution of class labels is detailed in Table~\ref{tab:labels}.
\begin{table}[htbp]
    \centering
    \begin{tabular}{cccccc}
    \toprule
        Stage & W & N1 & N2 & N3 & R \\
    \midrule
        Percentage & 33\% & 3\% & 37\% & 14\% & 12\% \\
    \bottomrule
    \end{tabular}
    \caption{Stage distribution of the NSRR dataset.}
    \label{tab:labels}
\end{table}

\begin{figure}[htbp]
    \centering
    \begin{minipage}[t]{0.48\linewidth}
        \includegraphics[width=\linewidth]{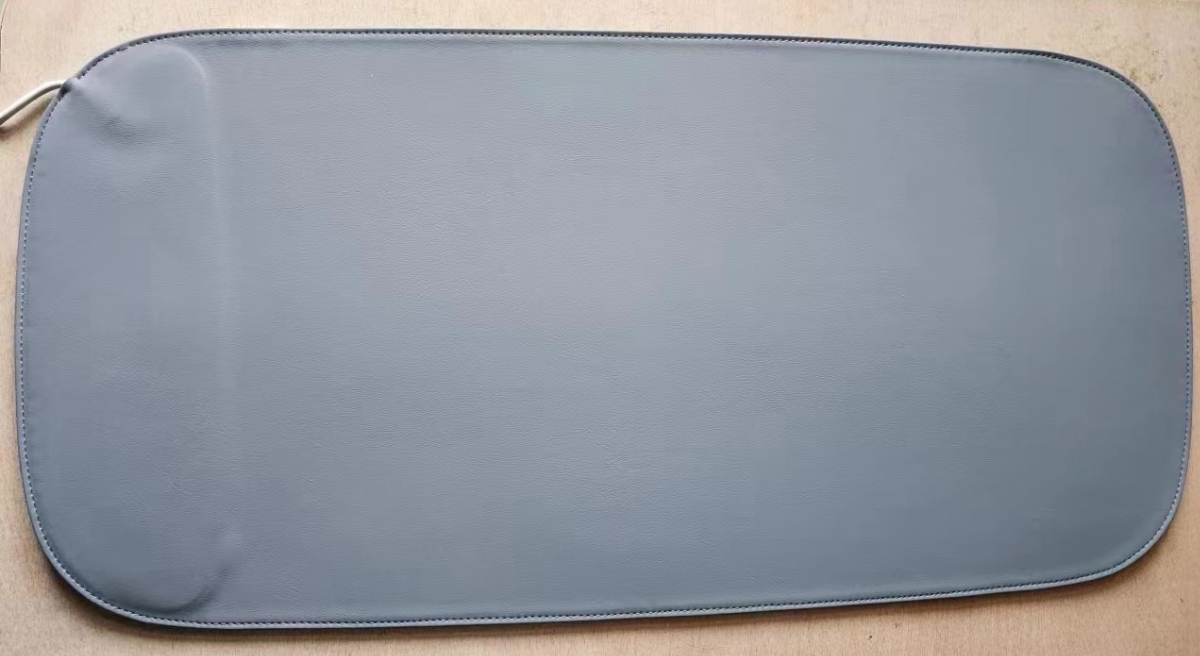}
        \caption{The prototype of the monitoring pad with the BCG sensor.}
        \label{fig:pad}
    \end{minipage}
    \hfill
    \begin{minipage}[t]{0.48\linewidth}
        \centering
        \includegraphics[width=\linewidth]{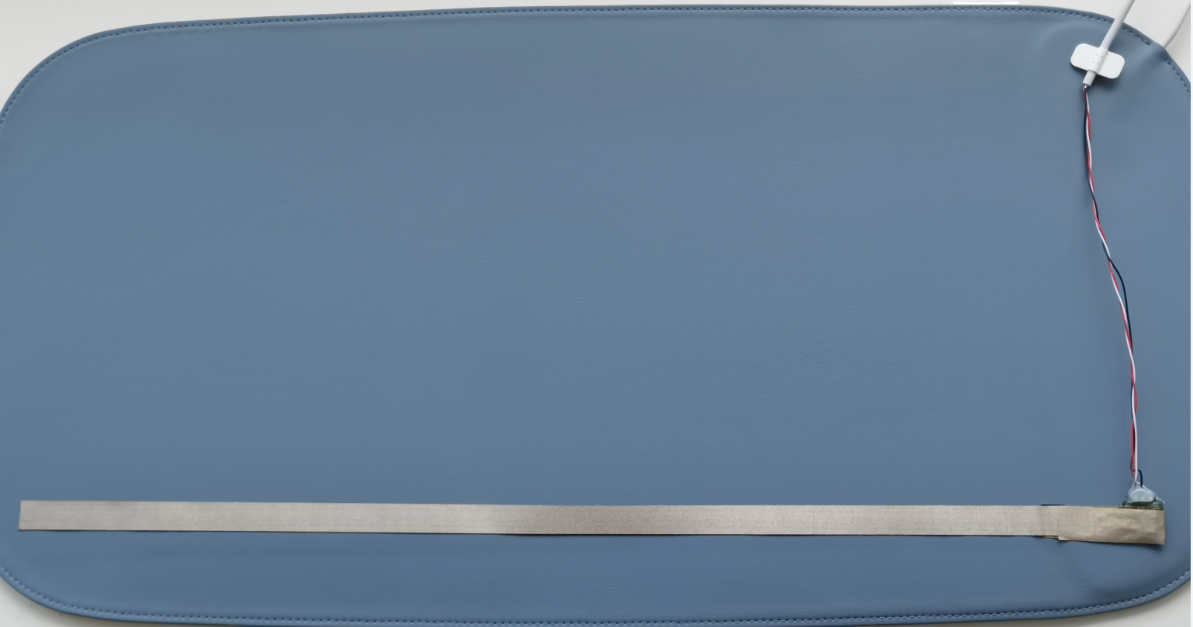}
        \caption{The monitoring pad and the unencapsulated piezoelectric sensors. The sensors are already placed in their related encapsulation positions.}
        \label{fig:pad-sensor}
    \end{minipage}
    \Description{A long strip of piezoelectric sensor is encapsulated in the monitoring pad. The sensor is placed near one of the long edges of the pad. A wire connects one end of the piezoelectric sensor with the external circuit to collect data.}
\end{figure}

\begin{table}[htbp]
    \centering
    \begin{tabular}[b]{rlc}
    \toprule
        AHI Group & Criterion & Percentage \\
    \midrule
        Normal & $\mathrm{AHI}<5$ & $18.4\%$ \\
        Mild Sleep Apnea & $5\le\mathrm{AHI}<15$ & $20.5\%$ \\
        Moderate Sleep Apnea & $15\le\mathrm{AHI}<30$ & $18.0\%$ \\
        Severe Sleep Apnea & $\mathrm{AHI}\ge 30$ & $43.1\%$ \\
    \bottomrule
    \end{tabular}
    \caption{The apnea-hypopnea index (AHI) distribution of the BCG265 dataset. Most (81.6\%) of the subjects suffer from sleep apnea, and 43.1\% are severe.}
    \label{tab:BCG-ahi}

    \begin{minipage}[t]{0.48\linewidth}
        \centering
        \begin{tabular}[b]{cc}
        \toprule
            Age Group & Percentage \\
        \midrule
            0--20 & $1.4\%$ \\
            20--40 & $56.5\%$ \\
            40--60 & $33.2\%$ \\
            60+ & $8.8\%$ \\
        \bottomrule
        \end{tabular}
        \caption{The age distribution of the BCG265 dataset. Most (89.7\%) subjects are 20 to 60 years old. Given that most subjects are sleep apnea patients, this may be a joint effect of prevalence and visiting rate.}
        \label{tab:BCG-age}
    \end{minipage}
    \hfill
    \begin{minipage}[t]{0.48\linewidth}
        \centering
        \begin{tabular}[b]{cc}
        \toprule
            Sex & Percentage \\
        \midrule
            Male & $65.7\%$ \\
            Female & $34.3\%$ \\
        \bottomrule
        \end{tabular}
        \caption{The sex distribution of the BCG265 dataset. Males are more than females. This is because most subjects are sleep apnea patients. Male sleep apnea prevalence is higher than female.}
        \label{tab:BCG-sex}
    \end{minipage}
\end{table}

\paragraph{The BCG265 Dataset}
We also verify our method using a BCG dataset collected by our monitoring pad prototype. As shown in Fig.~\ref{fig:pad}, piezoelectric BCG sensors are encapsulated in the pad and can be placed under the pillow for sleep monitoring. The piezoelectric sensors are encapsulated near the lower edge of the monitoring pad, as shown in Figure~\ref{fig:pad-sensor}.

We provide further discussions of the BCG sensor in Appendix~\ref{sec:BCG-sensor}.

We deployed the monitoring pads in a hospital environment, recording BCG signals in synchronization with the PSG. Technicians in the hospital annotate the sleep stages according to the PSG, enabling us to obtain the ground truth sleep stages for the BCG signal.

We collected 265 parallel PSG/BCG records from 265 distinct individuals in the hospital, with one PSG and one BCG record for each patient. The collection has passed the ethical review. We will refer to the dataset by BCG265.

BCG265 consists of individuals with obstructive sleep apnea (OSA)-related diseases, with the specific apnea-hypopnea index (AHI) distribution shown in Table~\ref{tab:BCG-ahi}. The age and sex distribution is shown in Tables \ref{tab:BCG-age} and \ref{tab:BCG-sex}, respectively.
We provide the class label distribution of the BCG265 dataset in Appendix~\ref{app:data}.

The whole BCG265 dataset serves as the test dataset.

\subsection{Evaluation Metrics}

Sleep staging is a multi-class classification task. Therefore, we evaluate the proposed method by four main metrics, higher is better. Let $y_1, y_2, \ldots, y_T$ denote the ground truth sleep stages, and $\hat{y}_1, \hat{y}_2, \ldots, \hat{y}_T$ denote the predicted ones. The metrics are formulated as follows.

\textbf{Accuracy}: $\mathrm{ACC}=\frac{1}{T}\sum_{t=1}^T 1_{y_t=\hat{y}_t}$. In the multi-class classification setting, the micro F1 score equals $\mathrm{ACC}$.

\textbf{Cohen's $\kappa$}~\cite{cohen1960coefficient}: The coefficient takes the class imbalance into account:
\begin{displaymath}
    \kappa=\dfrac{T\sum_{t=1}^T 1_{y_t=\hat{y}_t}-\sum_{i=1}^T\sum_{j=1}^T 1_{y_i=\hat{y}_j}}{T^2-\sum_{i=1}^T\sum_{j=1}^T 1_{y_i=\hat{y}_j}}.
\end{displaymath}
Basically, $\kappa=1$ if $y$ and $\hat{y}$ are in perfect agreement, and $\kappa=0$ if $y$ and $\hat{y}$ are independent.

\textbf{Macro F1}: $\mathrm{MF1}=\frac{1}{C}\sum_{c=1}^C 2P_c R_c/(P_c + R_c)$, where $P_c$, $R_c$ are the $c^{\rm th}$ class precision and recall, respectively:
\begin{displaymath}
    P_c=\frac{\sum_{t=1}^T 1_{y_t=\hat{y}_t=c}}{\sum_{t=1}^T 1_{\hat{y}_t=c}}, R_c=\frac{\sum_{t=1}^T 1_{y_t=\hat{y}_t=c}}{\sum_{t=1}^T 1_{y_t=c}}
\end{displaymath}

\textbf{Weighted F1}: $\mathrm{WF1}=\sum_{c=1}^C w_c \cdot 2P_c R_c/(P_c + R_c)$, where $P_c$, $R_c$ are defined above, and $w_c=\frac{1}{T}\sum_{t=1}^T 1_{y_t=c}$.

\subsection{Training Details}
All experiments are repeated with 10 distinct seeds,
implemented in PyTorch~\cite{paszke2019pytorch}, and conducted on a single NVIDIA H100 80G GPU. Each model is trained by AdamW~\cite{loshchilov2017decoupled} for 50 epochs\ (300 steps each, 15k iterations in total), with a learning rate of $1.1 \times 10^{-4}$, a weight decay coefficient of $10^{-5}$, and a batch size of 32. The dropout probability of each Transformer encoder layer is set to 0.05.

The model has 29.5M trainable parameters. The data augmentation ensures the model with sufficient input samples to learn effectively from.

\section{Experimental Results}\label{sec:results}
This section presents the results of experiments conducted on four aspects: overall performance, generalization validation, baseline comparison, and ablation study.

\subsection{Overall Performance}

Table~\ref{tab:rq1} presents the comprehensive evaluation results of our SleepNetZero model for sleep staging, assessed on both the NSRR test split and the BCG265 dataset.
The results show several important aspects of the model's performance across different metrics and datasets:
\begin{table}[t]
\caption{The results of our SleepNetZero for sleep staging on the NSRR test split and the BCG265 dataset~(mean $\pm$ standard). The model is trained on NSRR. In the second column, the model is evaluated on the NSRR test split. In the third column, the model is evaluated on the BCG265 dataset. We observe a significant performance drop on BCG265, which can be considered a consequence of the sensor gap.}
\label{tab:rq1}
\centering
\begin{tabular}{ccc}
\toprule
Metric  & NSRR test split & BCG265 \\
\midrule
ACC     & $0.803 \pm 0.002$ & $0.699 \pm 0.007$ \\
$\kappa$& $0.718 \pm 0.003$ & $0.588 \pm 0.013$ \\
MF1     & $0.665 \pm 0.004$ & $0.612 \pm 0.009$ \\ 
WF1     & $0.792 \pm 0.001$ & $0.664 \pm 0.007$ \\
\bottomrule
\end{tabular}
\end{table}

\textbf{Accuracy~(ACC)}: The model achieves an accuracy of $0.803 \pm 0.002$ on the test split, indicating high overall correctness in its predictions. However, when evaluated on the BCG265 dataset, the accuracy drops to $0.697 \pm 0.007$. This decline could be attributed to the sensor gap with the lower quality of the BCG265 dataset, which is unseen in the training data.

\textbf{Cohen’s Kappa~($\kappa$)}: With a kappa score of $0.718 \pm 0.001$ on the test split and $0.589 \pm 0.011$ on the BCG265 dataset, the model demonstrates substantial agreement with the ground truth labels, albeit with a noticeable decrease on the BCG265 dataset. This decrease further confirms the challenges posed by the sensor gap introduced by the BCG265 dataset.

\textbf{Macro F1 Score~(MF1) and Weighted F1 Score (WF1)}: The MF1 and WF1 scores further elucidate the model’s capabilities in handling imbalanced data. Specifically, the MF1 scores of $0.664 \pm 0.005$ on the test split and $0.611 \pm 0.010$ on the BCG265 dataset, indicate room for improvement in achieving consistent performance across all classes.

The decline in performance on the BCG265 dataset can primarily be attributed to the sensor gap between the BCG and PSG signals.
It is difficult for a model primarily trained on PSG-aligned data to perform equally effectively on BCG data, as the model's feature detectors are optimized for signals of higher consistency and quality inherent to PSG.
Despite that, our method has made significant strides in addressing the decline resulting from sensor gaps. 
However, the ongoing performance discrepancies between datasets highlight that matching PSG signal accuracy remains challenging.

Additionally, we report confusion matrices of our model on both the NSRR test split and BCG265. Table~\ref{tab:first_table} shows the confusion matrix on the NSRR test split, while Table~\ref{tab:second_table} shows the confusion matrix BCG265. Each confusion matrix is the average of ten random runs and is normalized along the row direction.
Our method yields excellent results on the W, N2, and R stages, yet it struggles to accurately identify the N1 stages in the majority of instances. This issue can be attributed to two principal factors. First, the N1 stage exhibits the lowest annotation consistency among technicians, adding complexity to its identification~\cite{magalang2013agreement}. Second, the extreme imbalance of classes in the dataset, with only about 3\% containing N1 labels, further exacerbates the problem. Therefore, enhancing the model's performance on unbalanced labels, especially with the challenge posed by the N1 stage, remains a significant task for future work.

\begin{table}[htbp]
\centering
\begin{minipage}{.45\linewidth}
\centering
\begin{tabular}{cccccc}
\toprule
 & W     & N1    & N2    & N3    & R   \\ \midrule
W                  & 0.900 & 0.012 & 0.070 & 0.002 & 0.016 \\
N1                 & 0.268 & 0.122 & 0.516 & 0.002 & 0.093 \\
N2                 & 0.041 & 0.012 & 0.844 & 0.067 & 0.036 \\
N3                 & 0.008 & 0.000 & 0.459 & 0.527 & 0.005 \\
R                  & 0.026 & 0.007 & 0.093 & 0.002 & 0.872 \\ \bottomrule
\end{tabular}
\caption{The confusion matrix of our SleepNetZero model on the NSRR test split. The model is trained on the NSRR train set. The row labels are the true stages, while the column labels are the predicted stages. The confusion matrix is normalized along the row direction.}
\label{tab:first_table}
\end{minipage}
\hfill
\begin{minipage}{.45\linewidth}
\centering
\begin{tabular}{cccccc}
\toprule
 & W     & N1    & N2    & N3    & R   \\ \midrule
W                  & 0.920 & 0.013 & 0.052 & 0.003 & 0.013 \\
N1                 & 0.322 & 0.105 & 0.494 & 0.005 & 0.075 \\
N2                 & 0.067 & 0.024 & 0.820 & 0.036 & 0.053 \\
N3                 & 0.022 & 0.002 & 0.465 & 0.496 & 0.015 \\
R                  & 0.059 & 0.019 & 0.127 & 0.003 & 0.793 \\ \bottomrule
\end{tabular}
\caption{The confusion matrix of our SleepNetZero model on the BCG265 dataset. The model is trained on the NSRR train set. The row labels are the true stages, while the column labels are the predicted stages. The confusion matrix is normalized along the row direction.}
\label{tab:second_table}
\end{minipage}
\label{tab:two_tables}
\end{table}

\subsection{Generalization of SleepNetZero}
The generalization capability of SleepNetZero is thoroughly validated across diverse datasets encompassing variations in age, gender, race, and Apnea-Hypopnea Index~(AHI). These factors are critical as they can significantly influence the physiological patterns associated with sleep, thereby impacting the accuracy and reliability of sleep stage classification. To show comprehensive results, we use the cross-validation method to obtain the performance on the whole dataset. 
\begin{figure}[htbp]
    \centering
    \subfloat[Age]{
        \label{fig:A}
        \includegraphics[width=0.3\textwidth]{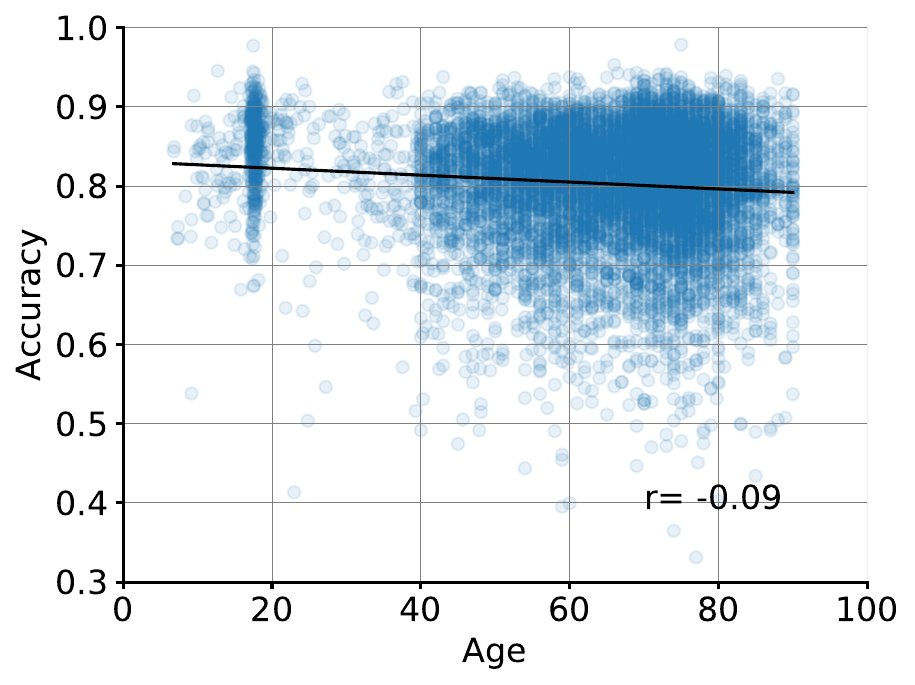}
    }
    \hfill
    \subfloat[BMI]{
        \label{fig:B}
        \includegraphics[width=0.3\textwidth]{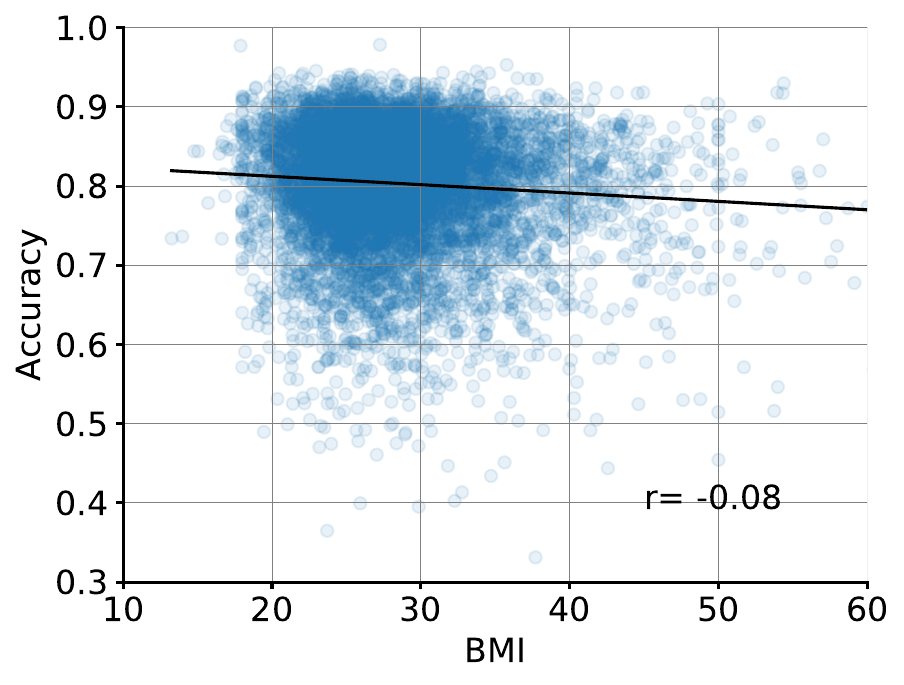}
    }
    \hfill
    \subfloat[Sex]{
        \label{fig:C}
        \includegraphics[width=0.3\textwidth]{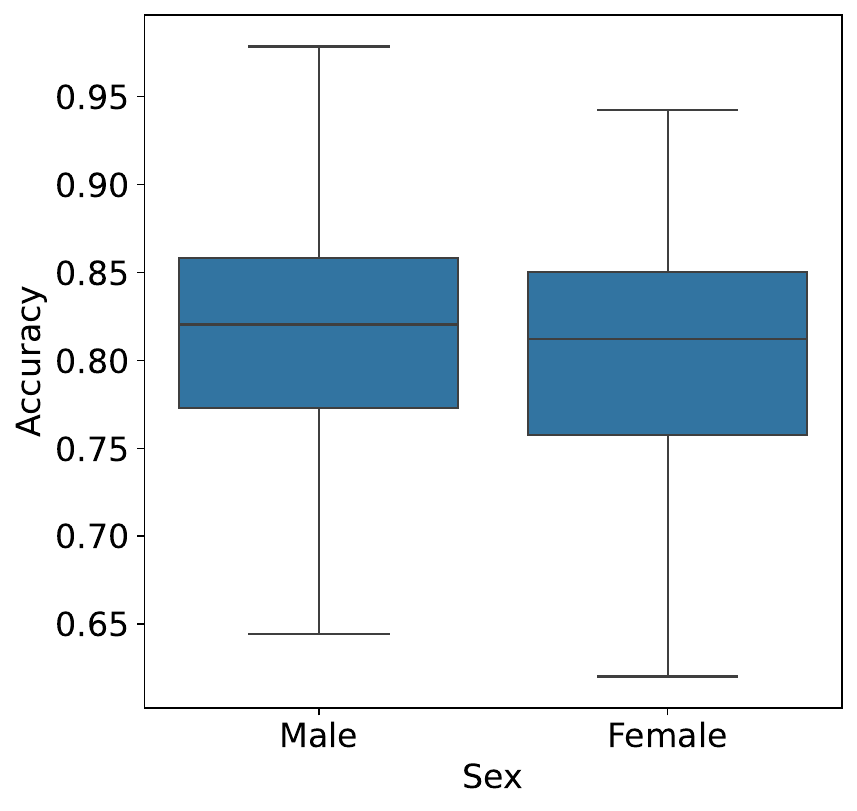}
    }

    \vspace{0cm}

    \subfloat[Transition Rate]{
        \label{fig:D}
        \includegraphics[width=0.3\textwidth]{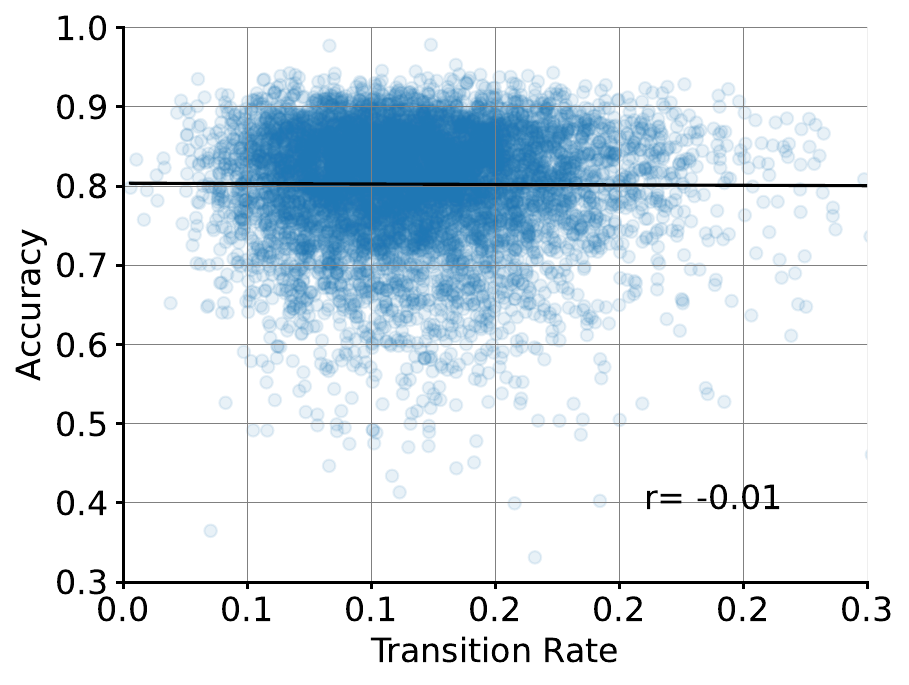}
    }
    \hfill
    \subfloat[AHI]{
        \label{fig:E}
        \includegraphics[width=0.3\textwidth]{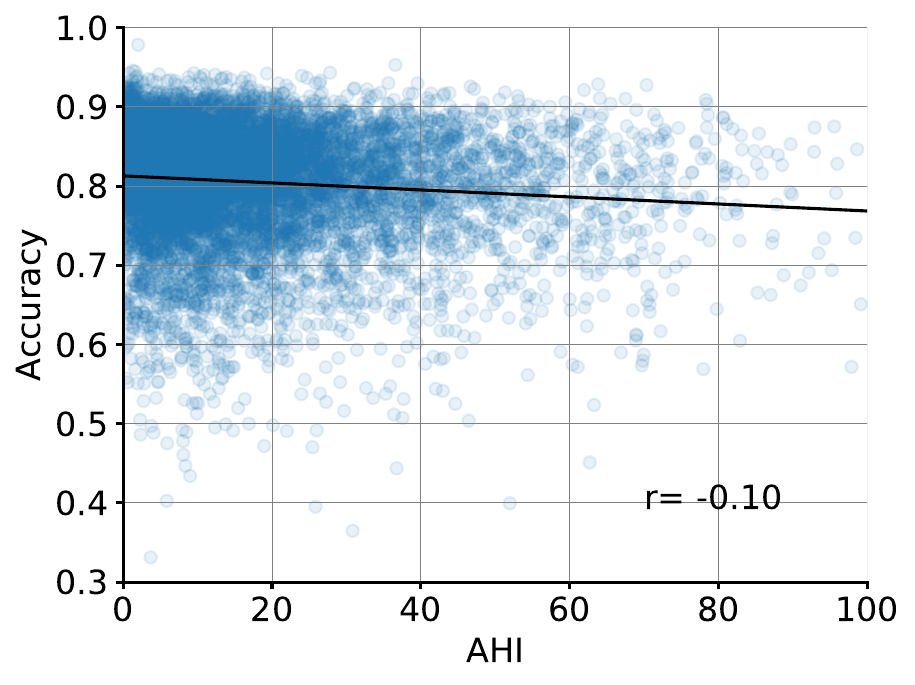}
    }
    \hfill
    \subfloat[Race/Ethnicity]{
        \label{fig:F}
        \includegraphics[width=0.3\textwidth]{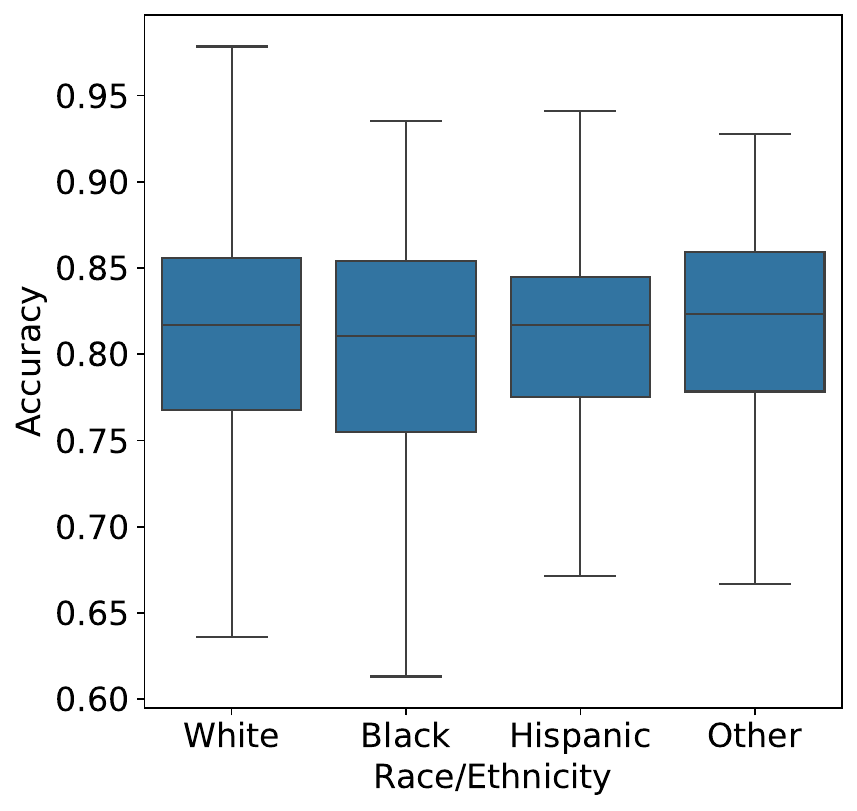}
    }
    \caption{Accuracy of the testing nights as a function of age, BMI, sex, transition rate, race, and AHI. The transition rate is defined as the ratio of transitions between different sleep stages to the total number of annotations. In sleep staging, an annotation is made every 30 seconds. If two consecutive annotations indicate different sleep stages, this is considered a transition. The transition rate is calculated by dividing the number of these transitions by the total number of 30-second annotations.}
    \label{fig:generalization}
    \Description{
        Six plots showing how accuracy changes with age, body mass index (BMI), sex, transition rate, apnea-hypopnea index (AHI), and race/ethnicity, respectively.
        In general, our proposed method performs slightly better for younger, thinner, male individuals, individuals with lower AHI, and individuals racial/ethnic groups other than white, black and hispanic. Transition rate does not affect the performance.
    }
\end{figure}

\subsubsection{Generalization across Age}:
The physiological signals associated with sleep stages can vary significantly with age. For instance, older adults may exhibit less pronounced sleep spindles, a feature crucial for identifying specific non-REM sleep stages. As shown in Figure~\ref{fig:A}, age exhibits a slight negative but significant correlation with accuracy~(r=-0.09, p<0.001). Although our dataset primarily consists of older adults, SleepNetZero demonstrates robust performance across all age groups, particularly in younger individuals. 

\subsubsection{Generalization across BMI}:
BMI~(Body Mass Index) characterizes a person's health in terms of weight and height. People with higher BMI tend to have a higher risk of sleep apnea and less percentage of deep sleep. As shown in Figure~\ref{fig:B}, BMI exhibits a slight negative but significant correlation with accuracy~(r=-0.08, p<0.001).

\subsubsection{Generalization across Gender}
Gender differences in sleep patterns, such as variations in REM sleep duration and sleep cycle architecture, have been documented in sleep research. As shown in Figure~\ref{fig:C}, SleepNetZero has a better performance on Male subjects than Female ones~(Welch’s T=7.98, p<0.001~\cite{welch1947generalization}).

\subsubsection{Generalization across Transition Rate}
The transition rate stands for the ratio of sleeping stages, as defined by the human scoring, that is different from the sleeping stage 30 seconds before. We use transition rate to represent the sleep staging stability of a sample. As shown in Figure~\ref{fig:D}, we found that it is not a significant determinant of accuracy~(r=-0.01, p=0.581).

\subsubsection{Generalization across AHI}
The Apnea-Hypopnea Index~(AHI) quantifies the severity of sleep apnea based on the number of apneas and hypopneas per hour of sleep. AHI categorizes the severity of sleep apnea into four levels: healthy~(\(AHI < 5\)), mild~(\(5 \leq AHI < 15\)), moderate~(\(15 \leq AHI < 30\)), and severe~(\(AHI \geq 30\)). These levels profoundly influence sleep architecture and quality.
As shown in Figure~\ref{fig:E}, AHI was significantly correlated with accuracy~(r=-0.10, p<0.001).

\subsubsection{Generalization across Race}
Racial and ethnic differences can influence sleep architecture and susceptibility to sleep disorders. We counted the accuracy rates on different racial groups including Caucasian~(White), African American~(Black), Hispanic, and other races~(Other).
As shown in Figure~\ref{fig:F}, we found that race can influence the accuracy significantly~(ANOVA~\cite{st1989analysis}, F=6.24, p<0.001).

\subsection{Comparison to Related Works}
We compare SleepNetZero with the related works in two aspects. First, we compare it with other BCG-based methods. Further, we compare it with the state-of-the-art sleep staging methods based on various signals.

\paragraph{BCG-based Sleep Staging}

As shown in Table~\ref{tab:cmp-bcg}, all the existing BCG-based methods were evaluated on restricted datasets (no more than 25 subjects). Therefore, SleepNetZero is superior in reliability, as it is evaluated on ten times more subjects. Moreover, considering the difference in evaluation methods, SleepNetZero offers comparable performance to less reliable ones (-0.1 ACC compared to 5-class methods, -0.15 $\kappa$ compared to 3-class methods).

\begin{table}[htbp]
    \caption{The comparison of various BCG-based sleep staging methods. The results evaluated on less than 100 subjects are considered less reliable, while those evaluated on at least 100 subjects are considered reliable. Some methods adopted 3-class or 4-class evaluation metrics. The 4-class evaluation merges N1 with N2, and the 3-class evaluation merges N1 and N2 with N3, reducing the task difficulty. Therefore, 3-class or 4-class metrics are higher than the 5-class ones we adopt. The ``$\dag$'' marked method trained and tested the model on overlapping subjects, deteriorating the reliability. SleepNetZero is the only reliable BCG-based method. Further, it exhibits comparable performance to less reliable ones.}
    \label{tab:cmp-bcg}
    \small
    \begin{sc}
    \centering
    \begin{tabular}{lll|cc|lll}
        \toprule
        Method & Evaluation Method & \#Test Subjects & ACC & $\kappa$ & Reliability & Zero-Shot \\
        \midrule
        QD-TVAM~\cite{migliorini2010automatic} & 3-class & 17 & 0.77 & 0.55 & $\times$ Low & $\times$ No \\
        \citet{mitsukura2020sleep} & 5-class & 25 & 0.80 & N/A & $\times$ Low & $\times$ No \\
        DeepSleep~\cite{rao2019deepsleep} & 4-class & 25 & 0.74 & N/A & $\times$ Low & $\times$ No \\
        \citet{yi2019non}$\dag$ & 3-class & 5 & 0.85 & 0.74 & $\times$ Low & $\times$ No \\
        SleepNetZero~(Ours) & 5-class & \textbf{265} & 0.70 & 0.59 & $\checkmark$ High & $\checkmark$ Yes \\
        \bottomrule
    \end{tabular}
    \end{sc}
\end{table}

\paragraph{Sleep Staging via Various Signals}

As shown in Table~\ref{tab:cmp-signal}, when tested on components extracted from PSGs, SleepNetZero offers comparable performance to previous works (-0.02 ACC/$\kappa$ relative to the 4-class SOTA PPG method). When tested on BCGs, SleepNetZero suffers from the sensor gap and exhibits a performance drop. However, the sensor gap is brought from the zero-shot experimental setting, and such a setting ensures the reliability of the result. Despite this, SleepNetZero performs the best among all reliable zero-burden methods. Moreover, it exhibits a comparable performance (-0.09 ACC/-0.11 $\kappa$ relative to 4-class radio frequency, -0.06 ACC/$\kappa$ relative to 4-class wrist PPGs) compared to the less reliable methods.
\begin{table}[htbp]
    \caption{The comparison of sleep staging methods via various signals. The results evaluated on less than 100 subjects are considered less reliable (marked by ``$\downarrow$''), while those evaluated on at least 100 subjects are considered reliable (marked by ``$\uparrow$''). The methods marked by ``*'' adopted the 4-class evaluation metrics. The 4-class evaluation merges N1 with N2, reducing the task difficulty. Therefore, 4-class metrics are higher than the 5-class ones that we adopt. Our method offers comparable performance on the NSRR test dataset. On the BCG265 dataset, despite the BCG performance drop caused by the sensor gap, BCG sensors cause zero burden, and the zero-shot experimental setting ensures the reliability of the result. Yet, considering the reliability, SleepNetZero performs the best among zero-burden methods.}
    \label{tab:cmp-signal}
    \small
    \begin{sc}
    \centering
    \begin{tabular}{lll|cc|lll}
        \toprule
        Method & Test Dataset~(\#Subjects) & Signal & ACC & $\kappa$ & Burden & Zero-Shot \\
        \midrule
        SleepTransformer~\cite{phan2022sleeptransformer} & SHHS~\cite{quan1997sleep,zhang2018national}~(1737 $\uparrow$) & EEG & 0.88 & 0.83 & $\times$ High & $\times$ No \\
        \hline
        SleepPPG-Net~\cite{kotzen2022sleepppg}* & CFS~\cite{redline1995familial,zhang2018national}~(320 $\uparrow$) & Finger PPG & 0.82 & 0.74 & $\times$ High & $\times$ No \\
        \hline
        SleepNetZero~(Ours) & NSRR~(963 $\uparrow$) & PSG & 0.80 & 0.72 & $\times$ High & $\times$ No \\
        \hline
        \citet{wu2023sleep}* & MIT-BIH~(18 $\downarrow$) & ECG & 0.76 & N/A & $\times$ High & $\times$ No \\
        \hline
        \citet{radha2021deep}* & Eindhoven~\cite{fonseca2017validation}~(60 $\downarrow$) & Wrist PPG & 0.76 & 0.65 & $\bigcirc$ Fair & $\times$ No \\
        \hline
        \citet{hong2022end}* & Private~(219 $\uparrow$) & Sound & 0.70 & 0.53 & $\checkmark$ Zero & $\times$ No \\
        \hline
        \citet{zhao2017learning}* & RF-Sleep~(25 $\downarrow$) & Radio Frequency & 0.79 & 0.70 & $\checkmark$ Zero & $\times$ No \\
        \hline
        \citet{van2023contactless}* & Private~(46 $\downarrow$) & Remote PPG & 0.68 & 0.49 & $\checkmark$ Zero & $\checkmark$ Yes \\
        \hline
        SleepNetZero~(Ours) & BCG265~(265 $\uparrow$) & BCG & 0.70 & 0.59 & $\checkmark$ Zero & $\checkmark$ Yes \\
        \bottomrule
    \end{tabular}
    \end{sc}
\end{table}

\subsection{Ablation Study}
To show the effectiveness of the component extraction module and generalization module, we conduct the ablation study.

\begin{table*}[t]
\caption{Ablation study: Kappa and Accuracy on the NSRR test split and BCG265. The models are trained with the NSRR train split. Columns 2--3 are evaluated on the NSRR test split, while columns 4--5 are evaluated on BCG265. The first row evaluates the full SleepNetZero, while the other rows evaluate SleepNetZero without one of the components. We can see that each component here is crucial for the good performance of SleepNetZero, especially for zero-shot generalization on BCG265.}
\label{tbl:ablation}
\vskip 0.05in
\begin{center}
\begin{small}
\begin{sc}
\begin{tabular}{lccccc}
\toprule
Method & NSRR Kappa & NSRR ACC & BCG265 Kappa & BCG265 ACC \\
\midrule
SleepNetZero & $0.718 \pm 0.003$ & $0.803\pm 0.002$ & $0.588\pm 0.013$&$0.699 \pm 0.007$\\
-- w/o Body Movement & $0.704\pm 0.002$ & $0.795\pm 0.001$&$0.540\pm 0.012$&$0.663\pm 0.010$\\
-- w/o Heartbeat & $0.689\pm 0.004$ & $0.786\pm 0.002$&$0.542\pm 0.015$&$0.666\pm 0.009$\\
-- w/o Breath & $0.668\pm 0.002$&$0.771\pm 0.001$&$0.496\pm 0.006$&$0.632\pm 0.006$ \\
-- w/o Random Amplification & $0.711\pm 0.001$&$0.799\pm 0.001$&$0.554\pm 0.008$&$0.671\pm 0.003$ \\
-- w/o Speed Perturbation & $0.700\pm 0.001$&$0.791\pm 0.001$&$0.533\pm 0.010$&$0.657\pm 0.005$ \\

\bottomrule
\end{tabular}
\end{sc}
\end{small}
\end{center}
\vskip -0.1in
\end{table*}

\subsubsection{Ablation Study of Component Extraction Module}

We conduct the ablation study of each component by replacing the component with uniform random noise.
As shown in Table~\ref{tbl:ablation}, SleepNetZero performs the best, with each of the three extracted components playing a crucial role.
SleepNetZero without body movement has a more significant drop on BCG265 than on the test dataset, highlighting the critical role of body movement in sleep staging, especially for generalization on BCG. As BCG and PSG sensors handle body movements differently, successful modeling of body movements can greatly help generalization from PSG data to BCG data.
SleepNetZero without heartbeat shows a moderate decline on both the test dataset and BCG265, underscoring the effectiveness of heartbeat in the sleep staging task. SleepNetZero without breath exhibits the greatest decline on both the test dataset and BCG265, confirming the vital importance of this channel for sleep staging. Combining the above results, we can get the results that all three channels are intrinsically linked, and our innovative application of body movement signaling is well suited to help improve generalizability.

\subsubsection{Ablation Study of Generalization Module} 
As shown in Table~\ref{tbl:ablation}, removing the random amplification or speed perturbation leads to a significant drop in performance compared to the full model, especially on BCG265. This indicates that both elements are crucial for sleep staging and cross-modality generalization. Due to differences in signal sources, PSG data and BCG data have different distributions, and our data augmentation methods successfully aid SleepNetZero in generalizing across modalities.

\section{Discussion}\label{sec:discuss}

\subsection{Superiority of SleepNetZero}
This work's strengths lie in its novel SleepNetZero framework, which, for the first time, proposes aligning BCG signals with PSG and achieving better performance than all other BCG-based methods for sleep staging. With this groundbreaking design, we address the scarcity of BCG samples and enhance the reliability of BCG-based sleep staging methods. Former BCG-based methods are heavily limited by the small size of the notated BCG dataset. In contrast, our method reaches a competitive result by leveraging large-scale publicly available PSG datasets.

Novel generalization techniques, such as data augmentation, demonstrate improved performance across sensors, ages, BMIs, sexes, health conditions, and races, as shown in \ref{sec:results}. Some research has shown that there exists a gap between populations. With these generalization techniques we successfully get across the gap and reach a good performance across populations.

Extensive experiments validate the framework's superiority and reliability, further supported by integration into a non-intrusive monitoring pad. By now, our algorithm has been running in several hospitals for some time and the response has been excellent.
This affirms the real-world applicability of the SleepNetZero framework compared to previous studies~\cite{migliorini2010automatic,yi2019non,rao2019deepsleep,mitsukura2020sleep,wu2023sleep}.

\subsection{Limitations}

The main limitations of our work are identified as two problems.

First, methods based on component extraction and alignment face a twofold ceiling.
The extraction pipelines still rely on rule-based methods.
While they circumvent the scarcity of annotated data, they are not perfectly accurate, limiting the model performance.
Moreover, the complicated physiological meanings residing in BCGs may not be well-exploited.
In such cases, utilizing the raw BCG signal may be preferable, as evidenced by the recent practice in the PPG field~\cite{kotzen2022sleepppg}.

Second, the leveraged datasets may introduce biases in the population.
While large-scale PSG datasets offer diverse populations, the subjects are recruited under certain conditions.
For example, SHHS~\cite{quan1997sleep,zhang2018national}~(5.8k subjects), MrOS~\cite{blackwell2011associations,zhang2018national}~(2.9k subjects), and MESA~\cite{chen2015racial,zhang2018national}~(2k subjects) contains only 40 years or older people. SHHS oversamples snorers.
Therefore, patients and the elderly account for a large population.
Removing the model's preference towards pathological or elder sleep patterns remains a future topic.

\subsection{Future Work}

Future studies may focus on two directions.

\textbf{Learning on sparsely annotated BCG datasets}:
Despite the scarcity of annotated BCG data, non-annotated data are convenient to collect with the monitoring pad.
The data can be recorded every night without extra effort through monitoring pads distributed to the research participants' homes.
Large-scale BCG datasets may facilitate self-supervised learning methods, which learn a representation of BCGs, encoding morphology and contextual information.
Fine-tuning a downstream task model based on the pre-trained encoding network requires much less annotation.

\textbf{Sleep disorder diagnosis}:
The significance of sleep staging resides in sleep disorder diagnosis.
The implication of sleep stages still requires experienced doctors to reveal, which causes burdens to patients.
In contrast, we wish to pave the way for comprehensive in-home sleep analysis, discovering potential disorders, providing the consumer with early warnings, and preventing worse consequences.

\section{Conclusion}\label{sec:conclusion}

This work presents a novel neural network framework named SleepNetZero, which achieve high-precision sleep staging by extracting certain physiological components to leverage cross-modality dataset, leading an effective paradigm for BCG-based sleep analysis. The powerful deep-learning model improves performance and generalizability, as the dataset grows in orders of magnitude. The work also incorporates novel data augmentation techniques, which help generalization across signals of different quality. Our work contributes to reliable in-home automated sleep analysis systems, which may unlock the potential of early diagnosis of sleep disorders, and enable longitudinal monitoring that helps evaluate the therapy effects.

\begin{acks}
This work is supported by the Ministry of Science and Technology of China STI2030-Major Projects (No. 2021ZD0201900, 2021ZD0201902).
\end{acks}

\bibliographystyle{ACM-Reference-Format}
\bibliography{ref_dltech,ref_med,ref_misc,ref_related}

\appendix

\section{The Physiological Meanings of the Sleep Stages}\label{sec:stages}
\begin{itemize}
    \item \textbf{Wake~(W)}: The state of being awake, which typically occurs at the start and end of the sleep record, as well as during brief awakening periods throughout the night.
    \item \textbf{Non-REM Stage 1~(N1)}: The lightest stage of sleep, often considered a transition phase between wakefulness and more profound sleep stages.
    \item \textbf{Non-REM Stage 2~(N2)}: A stage of sleep that represents a deeper sleep than N1 and accounts for the majority of sleep time. It is characterized by specific brain waveforms such as sleep spindles and K-complexes.
    \item \textbf{Non-REM Stage 3~(N3)}: The deepest stage of non-REM sleep, often referred to as slow-wave sleep due to the presence of high amplitude, low-frequency brain waves. It plays a crucial role in physical recovery and memory consolidation.
    \item \textbf{REM Sleep~(R)}: The stage of sleep associated with rapid eye movements, where most vivid dreaming occurs, and is linked to brain regions involved in learning and memory.
\end{itemize}

\section{More Dataset Information}\label{app:data}
The age and sex distributions of the NSRR dataset are listed in Tables~\ref{tab:NSRR-age} and \ref{tab:NSRR-sex}, respectively.
\begin{table}[htbp]
    \centering
    \begin{minipage}[t]{0.48\linewidth}
        \centering
        \begin{tabular}[b]{cc}
        \toprule
            Age Group & Percentage \\
        \midrule
            0--20 & $3.9\%$ \\
            20--40 & $1.6\%$ \\
            40--60 & $22.6\%$ \\
            60--80 & $59.2\%$ \\
            80+ & $12.7\%$ \\
        \bottomrule
        \end{tabular}
        \caption{The age distribution of the NSRR dataset. Most ($71.9\%$) subjects are over 60 years old. This is because some of the datasets only enroll elderly people.}
        \label{tab:NSRR-age}
    \end{minipage}
    \hfill
    \begin{minipage}[t]{0.48\linewidth}
        \centering
        \begin{tabular}[b]{cc}
        \toprule
            Sex & Percentage \\
        \midrule
            Male & $60.3\%$ \\
            Female & $39.6\%$ \\
        \bottomrule
        \end{tabular}
        \caption{The sex distribution of the NSRR dataset. Males are more than females.}
        \label{tab:NSRR-sex}
    \end{minipage}
\end{table}

In the BCG265 dataset, the distribution of class labels is detailed in Table~\ref{tab:BCG265-labels}.
\begin{table}[htbp]
    \centering
    \begin{tabular}{cccccc}
    \toprule
        Stage & W & N1 & N2 & N3 & R \\
    \midrule
        Percentage & 25\% & 12\% & 34\% & 15\% & 14\% \\
    \bottomrule
    \end{tabular}
    \caption{Stage distribution of the BCG265 dataset.}
    \label{tab:BCG265-labels}
\end{table}

\section{A Further Discussion on the BCG Sensor}\label{sec:BCG-sensor}
\begin{figure}[htbp]
    \centering
    \includegraphics[width=0.9\linewidth]{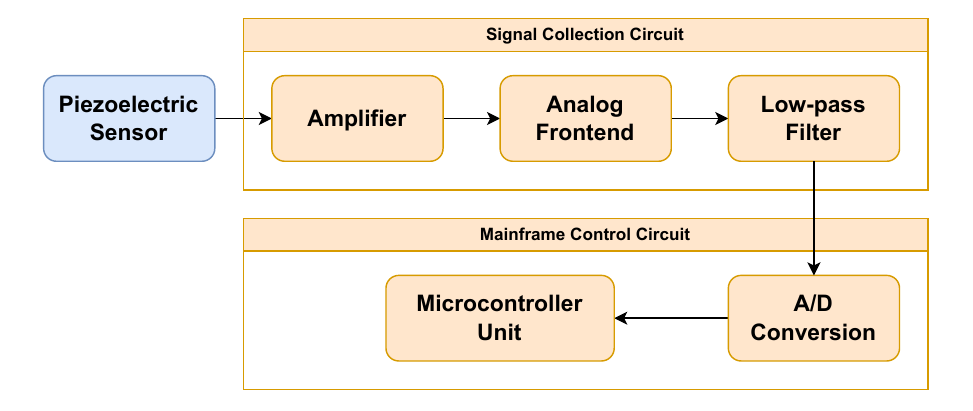}
    \caption{The system design of our monitoring pads.}\label{fig:sensor-sys}
    \Description{
        The signal from the piezoelectric sensor is fed into the signal collection circuit, and then processed by the mainframe control circuit.
        The signal collection circuit consists of three stages: the amplifier, the analog frontend, and the low-pass filter.
        The mainframe control circuit consists of two stages: A/D conversion and the microcontroller unit.
    }
\end{figure}

\paragraph{System Design}
The system design of our monitoring pads is shown in Figure~\ref{fig:sensor-sys}.

\paragraph{Comparison to other non-contact sensors}
In similar scenarios, sensors include optical fibers and piezoresistive sensors. Compared to piezoresistive sensors, our sensors have higher precision and sensitivity, and they are capable of capturing physiological information related to breathing and heartbeats. The accuracy and sensitivity of optical fiber sensors are somewhat higher than our piezoelectric sensors. However, they are more expensive, have complex encapsulation processes (requiring weaving into a mesh), and cannot withstand significant external pressure. Compared to these similar sensors, the piezoelectric sensors used in this paper meet the requirements for the precision and sensitivity of perceiving sleep activities and can still remain operational after external pressure.

\paragraph{Influence of co-sleepers}
In scenarios with co-sleepers, if the user themselves is also sleeping in the bed, the sensor's signal is mainly influenced by the user, with minimal impact from the co-sleeper. If the user is away from the bed, the sensor signal may be influenced by the co-sleeper, which is also one of the issues we need to address next. To obtain reliable data for testing the model, we currently require that only one person be present during the data collection process.

\section{Code Availability}
Source code is available at \href{https://github.com/nealchen2003/IMWUT2024-SleepNetZero}{https://github.com/nealchen2003/IMWUT2024-SleepNetZero}.

\end{document}